\documentstyle[prl,aps]{revtex}

\begin{document}

\draft

\title{Controlled Quantum Open Systems}

\author{Robert Alicki}

\address{Institute of Theoretical Physics and Astrophysics, University
of Gda\'nsk, Wita Stwosza 57, PL 80-952 Gda\'nsk, Poland}

\date{\today}
\maketitle

\begin{abstract}
The theory of controlled quantum open systems describes quantum systems interacting
with quantum environments and influenced by external forces varying according to given 
algorithms. It is aimed, for instance, to model quantum devices which can find applications 
in the future
technology based on quantum information processing. One of the main problems making difficult
the
practical implementations of quantum information theory is the fragility of quantum states
under external perturbations. The aim of this note is to present the 
relevant results concerning ergodic properties of open quantum systems which are useful for the
optimization of quantum devices and noise (errors) reduction. In particular we present
mathematical characterization of the so-called "decoherence-free subspaces" for discrete
and continuous-time quantum dynamical semigroups in terms of $C^*$-algebras and group
representations. We analyze the non-Markovian models also, presenting the formulas
for errors in the Born approximation. The obtained results are used to discuss the proposed
different strategies of error reduction.
\end{abstract}

\section{Introduction}
There exists a widespread believe that sooner or later the technology of the future will
employ typical quantum features of microscopic or mesoscopic systems like 
entanglement or superpositions of quantum states. The most prominent examples are the rapid
development of the quantum information and computation theory [1,2] and the corresponding progress
in experimental techniques. The later includes realization of quantum teleportation [3], 
monitoring of dissipation and decoherence for mesoscopic quantum systems [4], construction
of quantum gates [5], etc.
\par
The theory and experiment show that entanglement of separated systems and superpositions of
well distinguishable states are very sensitive to perturbations by environment. The
interaction with environment introduces noise (errors) which should be taken into account
by designing any quantum devices and their operations. The appropriate theory has been 
developed in the 70-ties and 80-ties as the Quantum Theory of Open Systems (see [6-11] for
the survey of this theory). The basic idea
of this theory is the decomposition of the Universe into the open system $S$ and the 
environment (reservoir , heat bath) $R$ with their corresponding Hilbert spaces ${\cal H}_S$
and ${\cal H}_R$ respectively. Assuming that the interaction between $S$ and $R$ is
weak we can always write, within a reasonable approximation, that for the initial moment 
$t_0 \equiv 0$ the initial state of the total system is given by the product formula
$$
\rho_{SR}(0) = \rho\otimes \omega_R
\eqno(1)
$$
where $\rho$ is an arbitrarily prepared initial state of $S$ and $\omega_R$ is a stable reference
state of $R$. Our interest is concentrated on the time evolution of $S$ for times $t \geq 0$ 
given by the reduced density matrix (we omit subscript $S$)
$$
\rho_t = {\rm Tr}_R\bigl( U_t \rho\otimes\omega_R U_t^*\bigr) = \Lambda_t(\rho)
\eqno(2)
$$
where $\{U_t ; t\in{\bf R}\}$ is a family of unitary operators acting on ${\cal H}_{SR} =
{\cal H}_S \otimes {\cal H}_R$ and describing the reversible (Hamiltonian) evolution
of the total system.  ${\rm Tr}_R$ is a partial trace over ${\cal H}_R$. 
The dynamical maps $\{\Lambda_t ; t\geq 0\}$ , their properties and different approximation
schemes for them (e.g. Markovian approximations) are the main topics of the Quantum Theory of 
Open Systems. 
\par
In order to deal with the problems posed by  quantum information processing and quantum 
computing one has to extend the standard approach introducing, beside the open system $S$
and the reservoir $R$, the third element -- the controler $C$. The role of $C$ is to prepare
an initial state , to measure a final one and to control, as far as it is possible, the time
evolution of $S$ by means of the time-dependent Hamiltonian $H_S(t)$. The presence of the 
time-dependent Hamiltonian makes, for example, the application of Markovian approximation 
often impossible and demands new mathematical techniques. Moreover, the physical structure of 
$C$, the resources used for state preparation and measurements must be taken into account in
the estimation of efficiency of quantum devices. We shall call such systems Controlled Quantum
Open Systems (CQOP).  
\par
One of the main problems in the theory of CQOP's is to reduce  
errors due to the interaction of a quantum device with an environment. At least three 
elements of the presented scheme can be optimized: the choice of the subspace in 
${\cal H}_S$ which supports the initial states of $S$, the trajectory $\{H_S (s) ; 0\leq s
\leq t\}$ which leads to a given unitary operation on ${\cal H}_S$ and the structure of 
the coupling of $S$ to $R$. In the recent literature one can find a number of proposals
of such optimization schemes [12-14]. However, some of these ideas are essentially based on 
the independent derivations of the earlier results, some other employ oversimplified 
phenomenological models which should be scrutinized from the point of view of a more 
fundamental theory. Therefore, it seems to be useful to collect, improve and generalize
the relevant results concerning mainly the ergodic properties of quantum open systems
and discuss their consequences for the optimization schemes mentioned above. 
\par
In the Section 2 we present mathematical characterization of the subalgebras of
observables which evolve according to reversible dynamics for a given irreversible
quantum dynamical map or quantum dynamical semigroup. Such subalgebras correspond to
"decoherence-free subspaces" in the language of modern quantum information theory.
The general ideas are illustrated by examples of collective dissipation models in Section3.
Section 4 is devoted to the derivation and discussion of the formula describing
errors in CQOP in the non-Markovian weak coupling Born approximation.

\section{Ergodic properties of quantum dynamical maps and semigroups }

In this Section we restrict ourselves in most cases to quantum open systems with finite-dimensional
Hilbert space ${\cal H}_S = {\bf C}^n$. We discuss the irreversible dynamics either in terms
of the single dynamical map $\Gamma $ and its iterations ${\Gamma^k , k=0,1,2,...}$ or the dynamical semigroup
$\{T_t ; t\geq 0\}$ with the composition law
$$
T_t T_s = T_{t+s}\ , t,s \geq 0\ ,\ T_0\equiv 1\ .
\eqno(3)
$$
Both $\Gamma$ and $T_t$ describe the dynamics in Heisenberg picture
i.e. they act on a matrix algebra ($C^*$-algebra) ${\cal M}_n$ representing quantum observables
of the system. In the following all
results independent on the semigroup property (3) are formulated in terms of a single
dynamical map $\Gamma$. The Schr\"odinger picture evolution
is given by the dual map $\Gamma^*$ acting on ${\cal M}_n$ which is treated now as the smallest complex linear space containing density matrices (quantum states) of the system and satisfying
$$
{\rm Tr}(A \Gamma(B))=  {\rm Tr}(\Gamma^*(A)B)\ .
\eqno(4)     
$$
Dynamical maps must preserve positivity and normalization i.e.
$$
\Gamma (A^* A) \geq 0\ ,\  \Gamma^*(B^* B)\geq 0\ ,
$$
$$
\Gamma ({\bf 1})= {\bf 1}\ , \ {\rm Tr}\Gamma^*(B) = {\rm Tr}(B)
\eqno(5)
$$ 
for all $A,B\in {\cal M}_n$.

\subsection{Complete positivity}

Positivity preserving condition (5) is generally to weak for a mathematically and 
physically consistent theory. The minimal
condition which must be imposed on dynamical maps to allow meaningful construction of joint
dynamics for noninteracting composed systems
is {\it complete positivity}. It means that for any $d = 1,2,3,...$, $\Gamma\otimes I_d$
is positive,where $I_d$ is an identity map acting on $d\times d$ matrices (i.e. 
trivial dynamical map on $d$-level quantum system). Any reduced dynamical map obtained
from the formula (2) is completely positive and any completely positive dynamical map
can be obtained from the reduced dynamics scheme. On the mathematical side completely positive 
maps on operator
algebras were studied already in the 50-ties and the celebrated 
Stinespring representation [15]
leads to a general form of completely positive dynamical map called often Kraus 
decomposition [16]
$$
\Gamma (A) = \sum_{\alpha}W_{\alpha}^* A W_{\alpha}
\eqno(6)
$$
where $W_{\alpha}$ are bounded operators on ${\cal H}$ satisfying 
$\sum_{\alpha}W_{\alpha}^*W_{\alpha} = {\bf 1}$. The decomposition (6) is highly nonunique,
in particular the sum over $\alpha$ can be replaced by an integral. If ${\cal H}$ is 
$n$-dimensional then complete positivity is equivalent to $n$-positivity and one can 
always find Kraus decomposition in terms of at most $n^2$ terms.
\par
From now on by the {\it quantum dynamical semigroup} (in the Heisenberg picture) we mean a  
a one-parameter strongly continuous semigroup ${T_t
 , t\geq 0}$ of  completely positive unity preserving maps. For finite dimensional case
we have always 
$$ 
A_t = T_t(A)= e^{tL} (A)\ ,\ {d\over dt} A_t = L (A_t)
\eqno(7)
$$ 
with the generator of the standard Lindblad, Gorini, Kossakowski and Sudarshan form [17] 
$$
L(A) = i[H, A] + \sum_{j=1}^p V_j^* A  V_j -{1\over 2}
\bigl\{\sum_{j=1}^p V_j^*V_j , A \bigr\}
$$
$$
= i[H, A] + {1\over 2}\sum_{j=1}^p\bigl( V_j^* [A , V_j] + [V_j^*, A]V_j \bigr)\ .
\eqno(8)
$$
The choice of the Hamiltonian $H=H^*$ and the operators (matrices) $V_j$ is again not unique and 
the sum over
$j$ can be replaced by an integral. In our finite dimensional case one can always find 
at most $p=n^2-1$ such $V_j$-s.
To simplify the notation we put always $\hbar\equiv 1$
and $k_B \equiv 1$ to have the same units for energy, frequency and temperature.
\par
The irreversible character of a dynamical map or semigroup is expressed in terms of
the following inequalities valid for any $A\in {\cal M}_n$
$$
\Gamma (A^* A)\geq \Gamma(A^*)\Gamma(A)\ --\ {\rm Kadison\ inequality}
\eqno(9)
$$
$$
L(A^* A)\geq L(A^*) A + A^* L(A)\ --\ {\rm dissipativity} .
\eqno(10)
$$
Indeed, puting equality in (9) we obtain in our case the reversible dynamical map
$A\mapsto U^* A U$ with a unitary $U$ while the equality in (10) implies the
Hamiltonian derivation structure $L(A)= i[H,A]$. Both inequalities are the consequences
of 2-positivity condition obviously satisfied for any completely positive dynamical
map and proved very useful in the analysis of ergodic properties.

\subsection {$C^*$-algebras and group representations}

In this Section we briefly review the basic mathematical facts about finite dimensional 
$C^*$-algebras and representation theory of compact groups which will be used later on [18].
\par
The full matrix algebra is an example of $C^*$- algebra i.e. a complex linear space with
adjoint operation (hermitian conjugation) $A\mapsto A^*$, multiplication $AB$ and complete 
with respect to the norm
satisfying $\|AA^*\|= \|A\|^2$. We are interested in finite-dimensional
$C^*$-algebras which can always be seen as $C^*$-subalgebras of full matrix algebras i.e.
linear subspaces of ${\cal M}_n$,
closed with respect to multiplication of matrices and hermitian conjugation and containing 
the unit matrix {\bf 1}. 
\par
Any such $C^*$-subalgebra ${\cal N}$ can be represented as the algebra of linear maps on
the finite dimensional Hilbert space in the following way. 
The Hilbert space is decomposed as
$$
{\bf C}^n = \bigoplus_{j=1}^m {\bf C}^{n_j}\otimes {\bf C}^{d_j}\ ,\ \ \sum_{j=1}^m n_jd_j = n
\eqno(11)
$$
where $n_j$ is the dimension of the irreducible representation of ${\cal N}$ and $d_j$ its
multiplicity. Then any element $A\in {\cal N}$ can be written in the form
$$
A =  \bigoplus_{j=1}^m A_j\otimes {\bf 1}_{d_j} 
\eqno(12)
$$
where $A_j \in {\cal M}_{n_j}$ and ${\bf 1}_{d_j}$ is a unit matrix acting on ${\bf C}^{d_j}$.
\par
In a general algebraic framework the physical symmetry or reversible dynamical map described in the Heisenberg picture
is given in terms of an {\it automorphism} of the corresponding $C^*$-algebra, i.e. a linear map $A\mapsto{\cal U}(A)$
satisfying ${\cal U}(A^*) = {\cal U}(A)^*$, ${\cal U}(AB) = {\cal U}(A){\cal U}(B)$ , 
${\cal U}({\bf 1})= {\bf 1}$ and reversible. For an automorphism of  a finite dimensional 
algebra ${\cal N}\subset {\cal M}_n$ 
reversibility follows from the former conditions and moreover there exists a unitary matrix
$U\in {\cal M}_n$ such that ${\cal U}(A) = U^* A U$. The group of all automorphisms
of ${\cal N}$ is denoted by $Aut({\cal N})$. A subgroup $Aut_0({\cal N})\subset
Aut({\cal N})$ contains all {\it inner automorphisms} given in terms of unitaries
$U\in {\cal N}$. For our finite dimensional case $Aut_0({\cal N})$ coincides with
the {\it path connected component of identity} of $Aut({\cal N})$. In order to illustrate the
fact, that even in the finite dimensional case not all automorphisms are inner, consider 
the abelian subalgebra of ${\cal M}_2$ of diagonal matrices and the "flip" map
$
\pmatrix{a & 0 \cr
         0 & b \cr} \mapsto \pmatrix{b & 0 \cr
                                   0 & a \cr}\ .
$
This automorphism can be writen as $A\mapsto \sigma_x A \sigma_x$, where the (unitary) Pauli matrix $\sigma_x$ is neither diagonal nor cannot be continuosly transformed into identity.

\par
We denote by ${\cal A}'= \{X\in {\cal M}_n ; [X,{\cal A}]=0\}$ the {\it commutant} of the subset
${\cal A}\subset{\cal M}_n$. If ${\cal A}$ is self-adjoint, i.e. $A\in{\cal A}\Rightarrow 
A^*\in{\cal A}$, then the commutant ${\cal A}'$ is a unital $C^*$-subalgebra of ${\cal M}_n$.
For a subalgebra ${\cal N}$ we have $({\cal N}')'={\cal N}$. Obviously if ${\cal A}\subset
{\cal B}$ then ${\cal B}'\subset {\cal A}'$. Denoting by $Alg({\cal A})$ the smallest $C^*$-
algebra containing ${\cal A}$ we have ${\cal A}' = Alg({\cal A})'$. For an algebra
${\cal N} = \bigoplus_{j=1}^p {\cal M}_{n_j}\otimes {\bf 1}_{dj}$ (see (12)) its commutant
${\cal N}' = \bigoplus_{j=1}^p {\bf 1}_{n_j}\otimes{\cal M}_{d_j}$.

\par
The theory of group representations provides natural examples of finite dimensional
$C^*$-algebras. Take a finite-dimensional unitary representation ${\cal R}: G\mapsto 
{\cal M}_n$; $g\mapsto R(g)$ of the compact group $G$ (discrete or Lie group). Such representation
can be always decomposed into irreducible representations ${\cal R}_j$ according
to (11). Any subspace ${\bf C}^{n_j}$ is a carrier of an irreducible representation
and appears $d_j$-times in the decomposition. Hence any unitary $R(g)\in {\cal R}(G)$
is decomposed as $R(g) =  \bigoplus_{j=1}^p R_j(g)\otimes {\bf 1}_{d_j}$. The $C^*$-algebra
generated by ${\cal R}(G)$ will be denoted by $Alg({\cal R}(G))$ and consists of elements
of the form $A = \sum_g \alpha_g R(g)\ ,\ \alpha_g\in{\bf C}$ which can be written as (12).

\subsection{Decoherence-free subalgebras }

For an open system with a time-independent Hamiltonian interacting with a stable environment 
being at the equilibrium state (heat bath) we expect that for an arbitrary initial state
the system returns to its thermal equilibrium state determined by the temperature of environment.
In mathematical terms it means that $\Gamma^*$ (or $T^*_t$) possesses a unique stationary state
$\rho_{eq}$ and $\lim_{n\to\infty}\Gamma^n(A) = {\rm Tr}(\rho_{eq} A){\bf 1}$ 
($\lim_{t\to\infty}T_t(A) = {\rm Tr}(\rho_{eq} A){\bf 1})$. We say that the dynamics $\Gamma$
($T_t$) is {\it ergodic} in this case. The relaxation process can be very complex and 
involving many
time scales of very different magnitude, some of them being much longer than the time scale 
relevant for the operation of our quantum device. Therefore, it is often very useful to study
simplified approximative nonergodic dynamics which take into account only the fastest
relaxation processes. For such a dynamics we are interested in observables and states which
evolve according to the reversible (unitary , Hamiltonian) evolution i.e. in a "decoherence-free" 
way. We shall prove that such observables are hermitian elements
of a certain unital $C^*$- subalgebra of ${\cal M}_n$. In practice, the manifest form of
this subalgebra is difficult to find and quite often we must be satisfied with a detailed knowledge
of its certain nontrivial subalgebras. It turns out that the symmetries of the system can be
very useful in this context. 
\par
The mathematical techniques used here are based on the earlier results on the ergodic theory
of quantum dynamical semigroups [19,20]. Some partial results of this type can be found in
the recent literature also [21].
\par
A minimal condition which should be satisfied by a decoherence-free observable $A$ with respect
to the dynamical map $\Gamma$ is the 
{\it lack of dissipation} expressed by
$$
\Gamma(A^2) = \bigl(\Gamma(A)\bigr)^2\ ,\  A=A^*\ .
\eqno(13)
$$
We formulate this condition in terms
of a {\it dissipation function} $D_{\Gamma}(.,.)$ from ${\cal M}_n \times {\cal M}_n$ to
${\cal M}_n$ defined by
$$
D_{\Gamma}(A,B) = \Gamma (A^* B) - \Gamma (A^*)\Gamma (B)
\eqno(14)
$$
Denote by ${\cal N}_{\bf R}(\Gamma)$ the following subset of hermitian matrices (observables)
in ${\cal M}_n$ and by ${\cal N}(\Gamma)$ its complexification
$$
{\cal N}_{\bf R}(\Gamma) = \{ A=A^*\in {\cal M}_n; D_{\Gamma} (A,A) =0 
\}\ ,
$$
$$
{\cal N}(\Gamma)=\{ A+iB ; A,B\in{\cal N}_{\bf R}(\Gamma)\}\ .
\eqno(15)
$$ 
We shall call ${\cal N}(\Gamma)$ the {\it decoherence free subalgebra} of $\Gamma$ and this
name is justified by the following theorem
\par
{\bf Theorem 1}
\par
1) ${\cal N}(\Gamma)$ is a unital $C^*$-subalgebra of ${\cal M}_n$ 
\par
2)$\Gamma$ restricted to ${\cal N}(\Gamma)$ is a unity preserving homomorhism, i.e. 
for any $A , B\in {\cal N}(\Gamma)$
$$
\Gamma (AB) =  \Gamma(A)\Gamma(B) \ ,\ \Gamma ({\bf 1})={\bf 1}
\eqno(16)
$$
and moreover $\Gamma\ {\rm is \ invertible\ on}$
$\Gamma\bigl({\cal N}(\Gamma)\bigr)$.

3) There exists a unitary $U_{\Gamma}\in {\cal M}_n$ such that
$$
\Gamma (A) = U^*_{\Gamma} A U_{\Gamma}\ ,\ {\rm for\ all}\ A \in {\cal N}(\Gamma) .
\eqno(17)
$$
\par
{\bf Proof}
We show first that for any $A\in {\cal N}_{\bf R}(\Gamma)$
and arbitrary $B\in{\cal M}_n$
$$
\Gamma (AB)= \Gamma(A)\Gamma(B)\ ,\ \Gamma (BA)= \Gamma(B)\Gamma(A)\ .
\eqno(18)
$$
The relations (18) follow from the Kadison inequality (9) applied to $A +\lambda B$ with an arbitrary $\lambda\in{\bf C}$. Namely,
$$
0\leq D_{\Gamma}(A+\lambda B,A+\lambda B)
$$
$$
= \Gamma (A^2) + |\lambda|^2 \Gamma(B^*B)+\lambda \Gamma (AB)
+ {\bar\lambda} \Gamma (B^*A)
$$
$$- \Gamma(A)^2- |\lambda|^2 \Gamma(B^*)\Gamma(B)
- \lambda \Gamma (A)\Gamma(B) - {\bar\lambda} \Gamma (B^*)\Gamma(A)
$$
$$
= |\lambda|^2\bigl(\Gamma(B^* B)-\Gamma(B^*)\Gamma(B)\bigr) 
$$
$$
+ \lambda\bigl(\Gamma(AB)-\Gamma(A)\Gamma(B)\bigr)       
+ {\bar\lambda}\bigl(\Gamma(B^*A)-\Gamma(B^*)\Gamma(A)\bigr)\ .
\eqno(19)
$$       
Dividing both sides of the inequality (19) by $|\lambda|$ and taking $|\lambda|\to 0$
we have for any $z\in {\bf C}, |z|=1$
$$
z\bigl(\Gamma(AB)-\Gamma(A)\Gamma(B)\bigr)       
+ {\bar z}\bigl(\Gamma(B^*A)-\Gamma(B^*)\Gamma(A)\bigr)\geq 0
\eqno(20)
$$       
what implies (18).
\par
It is now a simple exercise to prove using (18) that if $\ A,B\in {\cal N}(\Gamma)$, ${\rm then}\ A^*$,$ B^*$ , $A+B$, $AB\ \in {\cal N}(\Gamma)$.
Hence ${\cal N}(\Gamma)$ is a unital $C^*$-subalgebra and from the general results on finite dimensional $C^*$-algebras reviewed in Section 2.2 it follows that
(17) holds also.
\par
Theorem 1 gives us the general mathematical structure of observables
evolving in a reversible way under a single step dynamics. The physical meaning of the algebra ${\cal N}(\Gamma)$ is illustrated by the
most elementary physical process consisting of the following stages:
\par
1) preparation of the initial state $\rho$, 
\par
2) irreversible evolution governed by the dynamical map $\Gamma$,
\par 
3) measurement of a given observable with a spectral resolution
$$
A=\sum_k a_k P_k\ .
$$
\par
The probability of obtaining the value $a_k$ is equal to $ p_k = {\rm Tr}(\rho \Gamma(P_k))$
and if $A\in{\cal N}(\Gamma)$ then $p_k = {\rm Tr}(\rho U_{\Gamma}^*P_k U_{\Gamma})$.
In the language of quantum information theory it means that if we  choose decoherence-free
{\it decoding observables} then the noisy quantum channel is equivalent to a certain
noiseless one. 
\par
Analogical structures can be defined for the discrete (continuous) time dynamical semigroups
$\{\Gamma^k , k= 0,1,2,...\}$ ($\{T_t ,t\geq 0\}$) to obtain the global decoherence-free subalgebras which contain observables evolving in a reversible way for all times
$$
{\cal N}(\Gamma^{(\cdot)})= \bigcap_{k=1}^{\infty} {\cal N}(\Gamma^k)\ ,\ 
{\cal N}(T_{(\cdot)})= \bigcap_{t\geq 0} {\cal N}(T_t) 
\eqno(21)
$$
respectively.
\par
As a direct consequence of this definition, the Theorem 1 and the general results of 
Section 2.2 we obtain
\par 
{\bf Collorary} ${\cal N}(\Gamma^{(\cdot)})$ and ${\cal N}(T_{(\cdot)})$ are globaly invariant
under the action of $\Gamma^k $ and $T_t $ respectively. The both semigroups restricted
to decoherence-free algebras produce groups of unitary automorhisms $\{U_{\Gamma}^k ,
k\in {\bf Z}\}$ and $\{U_t ; t\in{\bf R}\}$. 
Moreover if for the discrete case
$U_{\Gamma}$ can be continuosly transformed into identity then $U_{\Gamma}
\in {\cal N}(\Gamma^{(\cdot)})$. Analogical result holds automatically for continuous
time semigroup, i.e. $U_t\in {\cal N}(T^{(\cdot)})$. 

\par
For the future applications it is useful to define a set of {\it fixed points} of $\Gamma$
as ${\cal F}(\Gamma)=\{A\in {\cal M}_n ; \Gamma(A) = A\}$. In general, ${\cal F}(\Gamma)$
is a self-adjoint linear subspace of ${\cal M}_n$ but under additional assumptions it
becomes $C^*$-algebra. For the dynamical semigroup ${\cal F}(T_{(\cdot)}) = Ker (L)=
\{A\in{\cal M}_n ; L(A)=0\}$.
\par
The algebras ${\cal N}(\Gamma)$, ${\cal N}(\Gamma^{(\cdot)})$ and 
${\cal N}(T_{(\cdot)})$ are the largest ones with the
decoherence-free property but their definition is not constructive and often difficult 
to apply. Therefore, it is sometimes convenient to consider smaller subalgebras which are 
easier to construct explicitely. Theorems 2,3  provide us with
the appropriate examples.
\par
{\bf Theorem 2}
\par
Assume that $\Gamma = {\cal U}\Gamma_D $ where
${\cal U}(X) = U^*XU$ is a unitary automorphism and
$$
\Gamma_D(X) = \sum_{\alpha =1}^r W_{\alpha}^* X W_{\alpha}\ .
\eqno(22)
$$
Then the commutants defined below are $C^*$-subalgebras and
$$
{\cal W}_1 = \{W_{\alpha}, W_{\alpha}^*; \alpha = 1,2,...,r\}' \subset {\cal W}_2 =
\{W_{\alpha}W_{\beta}^*;\alpha,\beta = 1,2,...,r\}'\subset {\cal N}(\Gamma)\ .
\eqno(23)
$$
If additionally ${\cal U}$ commutes with $\Gamma_D$ then
$$
{\cal W}_1\subset {\cal W}_3 = \{W_{\alpha}W_{\beta} , W_{\alpha}W_{\beta}^* , 
W_{\alpha}^*W_{\beta}^*;\alpha,\beta=1,2,...,r\}'\subset {\cal N}(\Gamma^{(\cdot)})\ .
\eqno(24)
$$   
\par
For continuous time dynamical semigroups we have:
\par
{\bf Theorem 3}
\par
Assume that the semigroup generator (8) decomposed as $L = L_H +L_D , L_H = i[H,\cdot]$ 
satisfies
$$
L_H L_D = L_D L_H\ .
$$
Then the commutant defined below is a $C^*$-subalgebra and
$$
\{V_j , V_j^*;j=1,2,...,p\}'\subset {\cal N}(T_{(\cdot)})\ .
$$
\par
{\bf Proof of Theorems 2 and 3}
From the definition of the commutant ${\cal W}_1
\subset {\cal W}_2 \subset {\cal N}(\Gamma_D)= {\cal N}(\Gamma)$.
\par
Due to the condition ${\cal U}\Gamma_D = \Gamma_D {\cal U}$, 
${\cal N}(\Gamma_D^{(\cdot)}) = {\cal N}(\Gamma^{(\cdot)})$. Finally, 
if $X=X^*$ commutes with all $W_{\alpha}W_{\beta}$ , $ W_{\alpha}W_{\beta}^*$ , 
$W_{\alpha}^*W_{\beta}^*, \alpha,\beta = 1,2,...,r$ then
$\Gamma^k(X^2) = \bigl(\Gamma^k (X)\bigr)^2 $ for all $k=0,1,2,...$
\par
Similarly, for continuous time semigroups the commutation of the Hamiltonian part $L_H$ with $L_D$ implies that the elements of the commutant $\{V_j ,V_j;j=1,2,...,p\}'$, which lie in a kernel of $L_D$, are decoherence-free.
\par
{\bf Remark} The decomposition of the dynamical map
into reversible and dissipative parts in Theorems 2,3 is not 
unique and can be optimized to obtain the largest commutants.
\par
To apply Theorems 2, 3 it is often convenient to use the results of the theory of group
representations. Assume that the
matrices $W_{\alpha}$ belong to $Alg({\cal R}(G))\subset {\cal M}_n$ of a certain unitary
representation of a finite or compact Lie group $G$.
In this case the Hilbert space is decomposed into carriers of irreducible representations
of $G$ as in (11) where $n_j$ are dimensions of irreducible representations and $d_j$ their 
multiplicities. The commutant 
$$
\{{\cal R}(G)\}'= \bigoplus_{j=1}^m {\bf 1}_{n_j}\otimes{\cal M}_{d_j}\subset {\cal W}_1
\eqno(25)
$$ 
in this case.
\par
Similarly, for quantum dynamical semigroup satisfying the assumptions of the Theorem 3
the matrices $V_j , V_j^*, j=1,2,...,p$ generate a certain Lie algebra in ${\cal M}_n$ and hence
the representation ${\cal R}(G)$ of the corresponding group $G$. In this case
the commutant (25) coincides with $\{V_j , V_j^*;j=1,2,...,p\}'$.
\par
Another application of the group theory is possible when the finite or compact Lie group
$G$ is a {\it symmetry group for the dynamical map} $\Gamma$ i.e.
$$
{\cal R}_g \Gamma {\cal R}_g^{-1} = \Gamma\ ,\ {\cal R}_g (X) = R(g)^* X R(g)
\eqno(26)
$$
for any unitary $R(g)\in{\cal R}(G), g\in G$. We call the condition (26) {\it global invariance} of the dynamical map $\Gamma$ with respect to the group $G$.
Using the definition of ${\cal N}(\Gamma^{(\cdot)})$
and then applying
(26) we can prove that from $A\in {\cal N}(\Gamma^{(\cdot)})$ it follows that ${\cal R}_g (A)\in {\cal N}(\Gamma^{(\cdot)})$ too. Therefore, the invariance holds 
$$
{\cal R}_g \bigl({\cal N}(\Gamma^{(\cdot)})\bigr) = {\cal N}(\Gamma^{(\cdot)})\ .
\eqno(27)
$$
If $G$ is a connected compact Lie group then any $R(g)$ can be continuosly connected to 
identity and hence by general theorems from Section 2.2 ${\cal R}_g$ is an inner automorhism
of ${\cal N}(\Gamma^{(\cdot)})$, i.e.
$R(g) \in {\cal N}(\Gamma^{(\cdot)})$. Hence the $C^*$-algebra generated by ${\cal R}(G)$ 
is included in
in the decoherence-free algebra ${\cal N}(\Gamma^{(\cdot)})$. The situation
for discrete symmetry may be quite different due to "incidental degeneracies" of $\Gamma$ 
which admit
larger group algebras. All that is true for the dynamical semigroup $\{T_t ; t\geq0\}$ also
and will be illustrated by examples in Section 3.  
\par
The stronger invariance conditions with respect to the symmetry group called {\it local invariance} can be formulated for 
the dynamical map $\Gamma ={\cal U}\Gamma_D
= \Gamma_D {\cal U}$ as in the Theorem 2. It reads for all $\alpha$ and  $g\in G$
$$
R(g)^*W_{\alpha} R(g) = W_{\alpha}
\eqno(28)
$$
Therefore the algebra generated by the group representation $Alg ({\cal R}(G))\subset 
{\cal N}(\Gamma^{(\cdot)})$. Similarly, for quantum dynamical semigroup as in the Theorem 3
the condition
$$
R(g)^*V_{\alpha} R(g) = V_{\alpha}
\eqno(29)
$$
implies $Alg ({\cal R}(G))\subset {\cal N}(T_{(\cdot)})= Ker (L_D)$.
\par
Summarizing, global invariance (26) for connected compact Lie group implies
$Alg ({\cal R}(G))\subset {\cal N}(\Gamma^{(\cdot)})$ while local invariance (28) or (29)
implies for any compact group $G$ that $Alg ({\cal R}(G))\subset {\cal N}(\Gamma^{(\cdot)})$ 
or  $Alg ({\cal R}(G))\subset {\cal N}(T_{(\cdot)})$ respectively.
\par  
The concrete examples of semigroups and their decoherence-free observables obtained using the
methods proposed above will be discussed in Section 3.

\subsection{ Limited relaxation}

We expect that the open system governed by a non-ergodic dynamics shows a behaviour
which can be called {\it limited relaxation}. It means that in the Heisenberg picture
any initial observable becomes for long times decoherence-free and the set of decoherence-free
observables is nontrivial. We are able to prove this property
under additional assumptions which are typically fulfilled for open quantum systems weakly
coupled to heat baths. 
\par
We begin with the dynamical map $\Gamma$ satisfying the following conditions:
\par
1) There exist a {\it faithful} (i.e. given by a strictly positive density matrix)
stationary state $\sigma$ , $\Gamma^*(\sigma) =\sigma$.
\par
2) {\it Detailed balance} condition with respect to $\sigma$ holds, i.e.
$\Gamma = {\cal U}\Gamma_D = \Gamma_D {\cal U}$ where
${\cal U}(X) = U^*XU$ is a unitary automorphism , ${\cal U}(\sigma) = \sigma$ and 
the dynamical map $\Gamma_D$ is a hermitian
operator on the Hilbert space $({\cal M}_n; <\cdot , \cdot>_{\sigma})$
equipped with the scalar product
$$
<A,B>_{\sigma} = {\rm Tr}(\sigma A^* B)\ ,\ A,B\in {\cal M}_n
\eqno(30)
$$
and the norm $\|A\|_{\sigma}= <A,A>^{1/2}$ . This Hilbert space is often called 
{\it Liouville space}. 
More explicitely ${\cal U}$ is unitary, $\Gamma_D$ is hermitian and  ${\cal U}\Gamma_D =
\Gamma_D{\cal U}$ is a {\it normal contraction} on the Liouville space$({\cal M}_n; <\cdot , \cdot>_{\sigma})$.

\par
The existence of a stationary state , generally non unique, is quaranteed by
the finite-dimensionality of the system's Hilbert space but its faithfulness is a nontrivial
assumption.
\par
{\bf Lemma 1}
\par
If the dynamical map $\Gamma$ possesses a faithful stationary state $\sigma$ then
the fixed point set ${\cal F}(\Gamma) $ is a $C^*$-subalgebra of ${\cal M}_n$ and  ${\cal F}(\Gamma)\subset
{\cal N}(\Gamma^{(\cdot)})$. 
\par
{\bf Proof} Take $A=A^*\in {\cal F}(\Gamma)$ then, $ A^2 = \bigl(\Gamma(A)\bigr)^2$ and hence
$$
0= {\rm Tr}\bigl( \sigma [A^2 - \Gamma(A)\Gamma(A)]\bigr) =
{\rm Tr}\bigl( \sigma (\Gamma(A^2) - \Gamma(A)\Gamma(A))\bigr) \ .
\eqno(31)
$$
As $\Gamma(A^2) - \Gamma(A)\Gamma(A)\geq 0$ and $\sigma$ is faithful then (31)
is equivalent to
$\Gamma(A^2)= \bigl(\Gamma(A)\bigr)^2$. It follows that $A\in{\cal N}(\Gamma)$.
If $A=A^*\in {\cal N}(\Gamma)$ then we can reverse the reasoning to prove that
$A\in {\cal F}(\Gamma)$. For a non-hermitian $A$ we can use the decomposition
$A= A_1 +iA_2$ into hermitian ones.
\par
We shall denote by $P_{\Gamma}$ the orthogonal projector acting on 
the Liouville space $({\cal M}_n; <\cdot,\cdot>_{\sigma})$ and projecting on the subspace
${\cal N}(\Gamma^{(\cdot)})$.
\par
Limited relaxation process is characterized by the following theorem.
\par
{\bf Theorem 4}
\par
Consider a dynamical map $\Gamma$ satisfying the detailed balance condition of
above. Then ${\cal N}(\Gamma^{(\cdot)}) = {\cal F}(\Gamma_D)$ and
$$
\lim_{k\to\infty}\|\Gamma^k (A) - {\cal U}^k(P_{\Gamma}A)\| = 0\ ,\ {\rm for\ all} 
\ A\in{\cal M}_n\ .
\eqno(32)
$$
\par
{\bf Proof}  From $\Gamma^k = {\cal U}^k\Gamma_D^k$ it follows that ${\cal N}(\Gamma^{(\cdot)})
={\cal N}(\Gamma_D^{(\cdot)})$. For any $A\in{\cal N}(\Gamma^{(\cdot)})$ we have
$\|\Gamma_D^k(A)\|_{\sigma} = \|A\|_{\sigma}$.Therefore, because $\Gamma_D$ is a hermitian
contraction on the Liouville space,  ${\cal N}(\Gamma_D^{(\cdot)})={\cal F}(\Gamma_D)$.

\par
For the dynamical semigroup with the generator $L= i[H,\cdot]+L_D$ the {\it detailed balance condition}
means that the Hamiltonian part $L_H$ commutes with $L_D$, $[H,\sigma] = 0$ and $L_D$
is a hermitian operator on the Liouville space $({\cal M}_n; <\cdot,\cdot>_{\sigma})$ [7].
Introducing a kernel ${\rm Ker}(L_D) = \{A\in{\cal M}_n; L_D(A)=0\}$, the orthogonal
projection on ${\rm Ker}(L_D)$ denoted by $P_{L_D}$ and applying the same idea of the proof
we obtain the following analog of the Theorem 4.
\par
{\bf Theorem 5} 
\par
Consider the dynamical semigroup $T_t$ with the generator satisfying detailed balance 
condition of above. Then ${\cal N}(T_{(\cdot)}) = {\rm Ker}(L_D)$ and
$$
\lim_{t\to\infty}\|T_t(A) - e^{itH}P_{L_D}(A)e^{-itH}\| = 0\ ,\ {\rm for\ all} 
\ A\in{\cal M}_n\ .
$$
\par
{\bf Example}
\par
The particular case of detailed balance generator with the stationary Gibbs state
$$
\sigma = Z^{-1}e^{-H/T}
\eqno(33)
$$
leads to a very specific structure of $L_D$ [7]
$$
L_D(A) =\sum_{j;\omega_j\geq 0}\Bigl( V_j^* A  V_j -{1\over 2}
\{V_j^*V_j , A \}+  e^{-\omega_j/T}\bigl(V_j A  V_j^* -{1\over 2}
\{V_jV_j^* , A \}\bigr)\Bigr)
\eqno(34)
$$
with $[H,V_j] = \omega_j V_j$ . Simple calculation involving the dissipativity condition (10)
and hermicity of $L_D$ on the Liouville space shows that in this case 
$$
0\leq {\rm Tr}\bigl(\sigma[L(A^*A) - L(A^*)A - A^*L(A)]\bigr)= - <A ,L_D(A)>_{\sigma}
$$
$$
= \sum_{j;\omega_j\geq 0}\Bigl(< [V_j, A], [ V_j, A]>_{\sigma}+ 
e^{-\omega_j/T} < [V_j^* , A], [V_j^* , A ]>_{\sigma}\Bigr)\ .
\eqno(35)
$$
Therefore, the
commutant $\{V_j,V_j^*;\omega_j \geq 0\}'= {\rm Ker}(L_D) ={\cal N}(T_{(\cdot)})$ .
 
\section {Examples}
In this Section we give examples of physical systems governed by nonergodic 
quantum dynamical semigroups and illustrate the different methods of searching
for decoherence-free observables introduced in Section 2.

\subsection {N-particle systems with permutation invariance}

Consider a system which consists of $N$ identical particles each of them is described
by a finite dimensional Hilbert space ${\cal H}^{(j)}\equiv {\bf C}^d ; j=1,2,...,N$.
There are two possibilities of constructing a quantum dynamical semigroup  generator
invariant with respect to the permutation group $S_N$. The first one corresponds to
particles interacting with identical "private" reservoirs and is given by the Master equation
in the Schr\"odinger picture for the $N$-particle density matrix $\rho_t$
$$
{d\over dt}\rho_t = -i [H ,\rho_t] + \sum_{m=1}^N L^*_m (\rho_t)
\eqno(36)
$$
where $H$ is a $N$- particle permutation invariant Hamiltonian and $L^*_m$ are identical copies
of a single-particle dynamical semigroup generator.
\par
The second example corresponds to the collective coupling of $N$-particle system to a single 
"common" reservoir 
$$
{d\over dt}\rho_t = -i [H ,\rho_t] + {1\over 2}\sum_{\alpha}\bigl([V_{\alpha},\rho_t V_{\alpha}]
 + [V_{\alpha}\rho_t, V_{\alpha}]\bigr)
\eqno(37)
$$
where $V_{\alpha}= \sum_{m=1}^N v^{(m)}_{\alpha}$ with identical copies of a single particle 
operator $v_{\alpha}$.
\par
The Hilbert space of the system possesses the structure of $N$-fold tensor product
$\bigotimes_N {\bf C}^d$ with the basis $\{e_{j_1,j_2,...j_N} = e_{j_1}\otimes e_{j_2}\otimes
\cdots\otimes e_{j_N}\}$. Any permutation $\pi\in S_N$ is represented by the operator
$R(\pi)$ defined in terms of the basis as
$$
R(\pi)e_{j_1,j_2,...j_N} = e_{\pi(j_1),\pi(j_2),...\pi(j_N)}\ .
\eqno(38)
$$
It is well-known that this representation is highly reducible and therefore 
$ Alg({\cal R}(S_N))$ is nontrivial. The explicite construction of the decomposition (11)(12)
is given in terms of Young tables and described in the texbooks on group theory [22].

\par
The semigroup generated by the master equation(36) is globally invariant with respect to $S_N$ but because
$S_N$ is a discrete group according to the results of Section 2.3 we can have no nontrivial
decoherence-free observables i.e. ${\cal N}(T_{(\cdot)})= {\bf C}$. Indeed, if a single-particle semigroup generated by $L^*_m$ is ergodic and the Hamiltonian $H=0$, then
the whole $N$-particle dynamics governed by (36) is ergodic too.  
\par
For the case of the dynamics goverened by the master equation (37) the situation is different. The corresponding semigroup is locally invariant with respect to $S_N$ and therefore according
to the results of Section 2.3 $ Alg({\cal R}(S_N))\subset {\cal N}(T_{(\cdot)})$.

\subsection{Superradiance model with $SU(2)$ symmetry} 

The superradiance [23] and subradiance phenomena [24] are the first examples of collective dissipative phenomena in quantum systems which leads to a nonergodic behaviour (on a certain
time scale) due to (approximative) permutation symmetry of the interaction of atoms with the electromagnetic field. In the simplest case of 2-level atoms in open space and for the temperature
$T=0$, we obtain the master equation of the form (37) with
$$
H= {1\over 2}\omega \sum_{m=1}^N \sigma_z^{(m)}\ ,\ v_{\alpha} = 
{1\over 2}\sqrt{\gamma}(\sigma_x + i\sigma_y)
\eqno(39)
$$
where $\omega$ is an atomic frequency, $\gamma$ is a radiation  damping constant and
$\sigma_j^{(m)}, j= x,y,z,$ are Pauli matrices for the $m$-th atom.
\par
In order to find decoherence-free observables we can use either the local symmetry of
the master equation with respect to the permutation group $S_N$ or the fact that the
collective operators $\sum_{m=1}^N \sigma_j^{(m)}$ which appear in the semigroup generator
define a reducible $N$-fold product representation of the group $SU(2)$ on $\bigotimes_N {\bf C}^2$.Then we obtain
$$
Alg ({\cal R}(S_N))\subset Alg({\cal R}(SU(2)))\subset {\cal N}(T_{(\cdot)})
\eqno(40)
$$
where ${\cal R}(S_N)$ is a standard representation of $S_N$ defined by (38) and 
${\cal R}(SU(2))$ is the mentiond above product representation of $SU(2)$. For our case of $T=0$ the explicite 
structure of
${\cal N}(T_{(\cdot)})$ is unknown while for $T>0$ we can use detailed balance condition and 
the Theorem 5 to characterize ${\cal N}(T_{(\cdot)})$ [20].

\section{Non-markovian controlled open systems}

The analysis of the previous Section has a rather mathematical and phenomenological
character and its most developed part is applicable to dynamical semigroups (discrete or
continuous) and hence involve Markovian assumption or at least stationary external
conditions. For a generic controlled system we expect that the external conditions
like, for instance, external electromagnetic fields used to control the quantum device 
vary in time. Therefore, we need
other approximation schemes which are based on the assumption of the {\it weak influence
of the environment on the system} and express the dynamics in terms of the fundamental
ingredients (controlled Hamiltonian, correlation functions of the bath)[25]. 

\subsection{ Errors in CQOP}

We would like to describe a quantum device operating during the time interval $[0,t]$.
We assume that in the initial moment the prepared state can be given by a product formula (1)
with a pure state $\rho = |\psi><\psi|$ while in the final moment the state of our open system is described by the reduced density
matrix (2).  The actual dynamics is compared with the ideal one given in terms of the controlled
unitary evolution $U_S(s)$ which solves the Schr\"odinger equation governed by the controlled
Hamiltonian $H_S(s) ; 0\leq s\leq t$. The error $\epsilon$ due to the interaction with environment can be defined as
$$
\epsilon = 1 - <U_S(t)\psi ,\Lambda_t (|\psi><\psi|)U_S(t)\psi>\ .
\eqno(41)
$$
The choice of the Hamiltonian $H_S(t)$ is a tricky problem. Typically, one takes a sum of the time-independent Hamiltonian $H_S^0$ of the bare system $S$ and a time-dependent contributions $H_S^1(s)$ describing the influence of controlled external fields on our system.
However the interaction with environment induces generally  time-dependent Hamiltonian corrections. For example, vacuum fluctuations of the electromagnetic field produce Lamb shift
of the atomic energy levels and van der Waals type interactions between neutral atoms [24].
Similar phenomena appear in solid state physics due to interactions with phonons and other quasi-particles. In the following we adopt the optimistic point of view assuming that we can include these corrections in the process of designing quantum algorithms. In practice one should expect
the presence of the Hamiltonian errors due to an approximative knowledge of the true 
physical Hamiltonian $H_S(s)$.

\subsection{Reduced dynamics in Born approximation}

We consider the system which consists of the quantum device $S$
driven by the time-dependent Hamiltonian $H_S(t)$ 
and the bath with its Hamiltonian $H_R$. Then the total Hamiltonian
reads 
$$
H(t)=H_S(t) +H_R + H_{int}
\eqno(42)
$$
with the interaction Hamiltonian of the form
$$
H_{int}=\sum_\alpha S_\alpha\otimes R_\alpha
\eqno(43)
$$
where $S_\alpha,R_\alpha$ are self-adjoint operators. 
\par
In the following we investigate the dynamical map $\Gamma^*$ in the Schr\"odinger
picture which describes the evolution of $S$ from the initial moment $t_{in}= -\tau$ to
to the final one $t_{fin} = \tau$ in Born approximation. The exact dynamical map can be 
expressed in terms of $H_{int}$, the unitary propagator for the total systems $U(t,s)$, 
the free
propagator for the total system $U_0(t,s)$ and the propagator $U_S(t,s)$ defined by
$$
U(t,s)={\bf T} e^{-i\int^t_sH(u)\; du}\ ,\ 
U_0(t,s)={\bf T} e^{-i\int^t_s(H_S(u)+H_R)\;du}\ ,
$$
$$
U_S(t,s)={\bf T} e^{-i\int^t_sH_S(u) \; du}
\eqno(44)
$$
where ${\bf T}$ is the time ordering operator. Introducing the "superoperator' notation:
${\hat U}\rho = U \rho U^*$ and ${\hat H}_{int}\rho = [H_{int},\rho]$,  choosing the initial
state similarly to (1)
$$
\rho_{SR}(-\tau)= \rho\otimes\omega_R\ ,\ [H_R,\omega_R]=0\ ,
{\rm Tr}(\omega_R R_{\alpha}) = 0
\eqno(45)
$$
and applying the second order integral identity for ${\hat U}(t,s)$ we obtain the exact equation 
$$
\Gamma^*(\rho)= {\hat U}_S(\tau,-\tau)\rho 
$$
$$
- {\rm Tr}\Bigl\{\int_{-\tau}^{\tau}
ds\int_{-\tau}^s du {\hat U}_0(\tau,s){\hat H}_{int} {\hat U}_0(s,u){\hat H}_{int} 
{\hat U}(u,-\tau)\rho\otimes\omega_R\Bigr\}\ .
\eqno(46)
$$ 
The Born approximation consists in replacing ${\hat U}(u,-\tau)\rho\otimes\omega_R$
by ${\hat U}_S(u,-\tau)\rho\otimes\omega_R$ in eq.(46). Our main assumption is that for
a given initial state $\rho$ and a given trajectory $\{H_S(t); -\tau \leq t\leq \tau\}$ 
of the controlling Hamiltonian the influence of the reservoir on the state of our system 
$S$ given by the second term on the RHS of eq.(46) is small, say of the order $\epsilon<<1$.  
Then up to the higher order terms $\sim \epsilon^2$ we can use the proposed approximation.
One should notice that we do not necessarily impose weak coupling or/and short time $\tau$
regime but we can search for the optimal choice of $\rho$ and $H_S(t)$ to fulfill
the Born approximation beyond this regime.
\par
The final formula for $\Gamma^*(\rho)$ in Born approximation reads
$$
\Gamma^*(\rho) = {\hat U}_S\Bigl(\rho + \Phi^* (\rho) - {1\over 2}\{K,\rho\}\Bigr)
\eqno(47)
$$
where ${\hat U}_S \equiv {\hat U}_S(\tau,-\tau)$.
The superoperator $\Phi^*$ called {\it error map} is a completely positive map given by
$$
\Phi^*(\rho) =\int_{-\tau}^{\tau}ds\int_{-\tau}^{\tau} du {\rm Tr}\bigl(\omega_R R_{\alpha}
R_{\beta}(s-u)\bigr) S_{\beta}(s,-\tau) \rho S_{\alpha}(u,-\tau)
\eqno(48)
$$
where $S_{\alpha}(s,u)= {\hat U}_S(u,s)S_{\alpha}$ , $R_{\alpha}(t)= e^{it{\hat H}_R}
R_{\alpha}$. The operator $K\geq 0$ is given by $K = \Phi ({\bf 1})$.
Strictly speaking one has also a contribution of the form $-i[h,\rho]$
which contains Hamiltonian corrections due to the interaction with $R$ like for example
Lamb shift, collective Lamb shift, etc. but we put $h\equiv 0$. This can be justified
by starting with a bare Hamiltonian $H^0_S(t)$ containing appropriate counterterms
and performing renormalization procedure. It is , however, equivalent to treating
$H_S (t)$ as a full physical Hamiltonian  and puting $h \equiv 0$. 
\par
The equation (47) resembles in its structure the short time approximation to
quantum dynamical semigroup and indeed, if the conditions of Markovian approximation
are fulfilled we can derive from (47) different types of the Markovian master equations [7,9,11].
The approximative map (47) is trace preserving but complete positivity is satisfied
up to the higher order terms in $\epsilon$. 
\par 
Finally, we pass to the frequency (energy) domain introducing spectral density
of the reservoir $R_{\alpha \beta}(\omega)$ by
$$
{\rm Tr}\bigl(\omega_R R_{\alpha}R_{\beta}(t)\bigr)=
\int_{-\infty}^{\infty} R_{\alpha\beta}(\omega)e^{-i\omega t}d\omega
\eqno(49)
$$
and defining
$$
Y_{\alpha}(\omega) = \int_{-\tau}^{\tau} S_{\alpha}(s,-\tau) e^{-i\omega s}ds\ .
\eqno(50)
$$ 
We obtain the following formula for the error map
$$
\Phi^*(\rho)= \sum_{\alpha,\beta}\int_{-\infty}^{\infty}d\omega\; R_{\alpha\beta}(\omega)\;
Y_{\beta}(\omega)\rho Y_{\alpha}^*(\omega)
\eqno(51)
$$
which is a kind of quantum {\it fluctuation-dissipation theorem} in the linear response regime
[9,10].

\subsection{Error formula}

The general definition of the error due to the interaction with the environment (41) can be 
applied to our approximative evolution
$$
\epsilon = 1- <U_S\psi , \Gamma^*(|\psi><\psi|)U_S\psi>\ .
\eqno(52)
$$
Putting the  expression (47) we obtain the error formula
in Born approximation
$$
\epsilon = <\psi, K \psi> - <\psi ,\Phi^*(|\psi><\psi|)\psi>
\eqno(53)
$$
which can be the starting point for the analysis of efficiency of quantum devices.
Writing it in terms of (49-51) we obtain
$$
\epsilon =\sum_{\alpha,\beta}\int_{-\infty}^{\infty}d\omega\; R_{\alpha\beta}(\omega)\;
\Bigl( <\psi ,Y_{\alpha}^*(\omega) Y_{\beta}(\omega)\psi> 
$$
$$
-
<\psi ,Y_{\alpha}^*(\omega)\psi><\psi, Y_{\beta}(\omega)\psi>\Bigr)\ .
\eqno(54)
$$
The interpretation of (54) becomes simple and elegant if we introduce the following
notation
$$
<\psi ,Y_{\alpha}^*(\omega) Y_{\beta}(\omega)\psi> -
<\psi ,Y_{\alpha}^*(\omega)\psi><\psi, Y_{\beta}(\omega)\psi>
= 2\tau S_{\alpha\beta}(\omega)\ .
\eqno(55)
$$
Now we can write the error formula in a compact form as an overlap
of two "correlators"  ${\cal R}(\omega) = [R_{\alpha \beta}(\omega)]$ and
${\cal S}(\omega) = [S_{\alpha \beta}(\omega)]$ multiplied by the duration
of the operation time of the quantum device $2\tau$
$$
\epsilon = 2\tau \int_{-\infty}^{\infty}d\omega\;{\rm Tr}\bigl({\cal R}(\omega)
{\cal S}(\omega)\bigr)\ .
\eqno(56)
$$
Both correlators encode the information about the dynamics and the initial state of 
the baths and
the quantum device, respectively. Moreover, one can see that their mathematical structure is
essentially the same. To show this we consider a sequence of random (quantum or classical)
time-dependent real variables $f_{\alpha}(t)$ and denote by $<\cdot>_F$ the appropriate average.
Define a correlator ${\cal F}(\omega) = [F_{\alpha\beta}(\omega)]$ by the following formula
$$
\lim_{\tau\to\infty}{1\over 2\tau} \int_{-\tau}^{\tau} ds\Bigl( <f_{\alpha}(t+s)
f_{\beta}(s)>_F -  <f_{\alpha}(t+s)>_F <f_{\beta}(s)>_F \Bigr)
$$
$$
=\int_{-\infty}^{\infty} F_{\alpha\beta}(\omega) e^{-i\omega t}d\omega\ .
\eqno(57)
$$
Then puting $f_{\alpha}\equiv R_{\alpha} , <\cdot>_F \equiv {\rm Tr}(\omega_R \cdot)$
we obtain ${\cal F}(\omega)\equiv {\cal R}(\omega)$. Similarly, 
${\cal F}(\omega)\equiv {\cal S}(\omega)$ for $f_{\alpha}
\equiv S_{\alpha} , <\cdot>_F \equiv <\psi ,\cdot \psi>$ if only $\tau$ is long enough
such that the limit $\tau\to\infty$ in (57) can be used.
\par
Having the formula (56) one can try to design different strategies reducing  errors. Although
very roughly the error seems to be proportional to the operation time $2\tau$ the situation
is much more complicated. For example, if the leading error source is described by the
correlator satisfying for small $\omega$
$$
{\cal R}(\omega)\sim \omega^{\kappa} \ ,\ \kappa >1
\eqno(58)
$$
the optimal strategy is to slow down the time evolution of $S$ in order to move the correlator 
${\cal S}(\omega)$ into the low frequency domain. On the contrary, if ${\cal R}(\omega)$
is flat "fast gates" are optimal. Generally, one can try to shift the support of 
${\cal S}(\omega)$ into the frequency domain where ${\cal R}(\omega)$ is small. However, because
both correlators are positively defined matrices we cannot expect dramatic reduction of errors
by some "cancelation effects". Moreover, different type of reservoirs and interactions produce
different shapes of ${\cal R}(\omega)$, e.g. scattering with gas particles gives flat shapes
while the linear coupling to quantum bosonic field produces the relations of type (58) with
some cut-offs for large $\omega$ (see [25]and the authors contribution to [11]). 
\par 
Another different strategy is to select an optimal subspace of initial states $\psi$
which are "decoherence-free" with respect to the leading error source. The following simple criterion is a generalization of the idea of quantum correcting codes [12].
\par 
Assume that the matrics ${\cal R}(\omega)$ is strictly positive for $\omega\in\Omega$.
Than any $\psi$ satisfying the eigenvector condition for all $\alpha$ and $\omega\in\Omega$
$$
Y_{\alpha}(\omega)\psi = \lambda_{\alpha}(\omega)\psi\ ,\ \lambda_{\alpha}(\omega)\in {\bf C}
\eqno(59)
$$
evolves without errors.
\par
Another partial results and examples of applications can be found in [25].

\section{Conclusions}

The author believes that the presented results, although often formulated in a rather abstract mathematical form, can serve as a guide in the discussion concerning the optimal strategies
in controlling quantum devices. These strategies can be roughly divided into passive and active
ones. At present, it seems that the passive ones can be realized in three apparently different ways:
\par
I) using the symmetries of the coupling to the baths and systems' Hamiltonian to obtain large enough decoherence-free subalgebras as presented in Section 2.3,
\par
II) slowing down the action of quantum algorithm (slow gates) for the environments which
satisfy the low frequency scaling (58), 
\par
III) Speeding up the quantum algorithm (fast gates) for the environments with flat ${\cal R}(\omega)\simeq {\cal R}$.
\par
The best known models of the type I) are given by the "superradiance" systems governed by
the equations (37). In principle, they can be realized in different settings using collective
coupling to bosonic fields (photons, phonons, etc.,). Such reservoirs satisfy also the condition
(58) what suggests that both I) and II) demand the similar technology. Indeed, the collective coupling of qubits to the bosonic field, which is a necessary condition for the validity of
(37) is possible if the wave-length of the relevant bosonic modes is long in comparison with the diameter of the device [24]. It means that only the low frequences can appear in the correlator
${\cal S}(\omega)$ (55),(56). This is equivalent to the slow gates strategy II). Moreover, in both
cases I) and II) the time scale of gates must grow with the size of the device (proportional
to the number of qubits) in order to keep the total error constant.
\par
The main obstacle for the strategy I) is the approximative character of the symmetry which becomes less accurate with the growing size of the device. Moreover, the Hamiltonian corrections due to the renormalization effects discussed in Section 4.2 do not possess this symmetry except
of the cases with the very special geometry [24].
\par
The strategies II) and III) are obviously contradictory  and therefore for real systems , for
which both types of reservoirs are at work, the optimization is needed.
\par
The most elaborated active strategy is the idea of quantum error corrections for quantum computers [12]. The existing models of error corrections possess, however, several drawbacks.
First of all, the noise is assumed to be independent of the algorithm what is equivalent
to the assumption of the strategy III) (${\cal R}(\omega)\simeq {\cal R}$). Moreover, the entropy produced in the computer must be very precisely taken away by "fresh" qubits prepared
in pure states. Finally, initial errors for all qubits used in this scheme are not taken into account seriously.
\par
Another active approach, called dynamical decoupling [13], involves the existence of
the frequency domain for which the correlator ${\cal R}(\omega)\simeq 0$. Then applying
a suitable fast modulation of the system's dynamics we can, in principle, move the support
of the correlator ${\cal S}(\omega)$ into this domain. Although for the specific environments
the correlator ${\cal R}(\omega)$ may have some "valleys", the overall tendency for ${\cal R}(\omega)$ is to grow with $\omega$. This is related to the fact that the number of accesible quantum states of the environment typically grows with increasing energy scale. In any case, dynamical decoupling strategy demands fast gates technology as for III).
\par
In authors' opinion the optimized combination of the passive strategies I,II and III may be the most realistic attempt to deal with quantum noise and associated errors. It would be interesting to apply the formulas of the type (56) to estimate the minimal error per a single gate within
the different proposed physical implementations of quantum computers (e.g. trapped ions , NMR , quantum dots , etc).
 
\acknowledgments
A part of this paper (Sections 4,5) is based on the collaboration with
Micha\l\ ,Pawe\l\ and Ryszard Horodecki.

\end{document}